\documentclass[12pt,preprint]{aastex}

\begin{document}

\title{On the Origin of Exponential Disks at High Redshift}

\author{Bruce G. Elmegreen \affil{IBM Research Division, T.J. Watson
Research Center, P.O. Box 218, Yorktown Heights, NY 10598, USA,
bge@watson.ibm.com} }
\author{Debra Meloy Elmegreen, David R. Vollbach, Ellen R. Foster, and Thomas E. Ferguson
\affil{Vassar College, Dept. of Physics \& Astronomy, Box 745,
Poughkeepsie, NY 12604; elmegreen@vassar.edu,
davollbach@vassar.edu, elfoster@vassar.edu, thferguson@vassar.edu}
}

\begin{abstract}
The major axis and ellipse-fit intensity profiles of spiral
galaxies larger than $0.3^{\prime\prime}$ in the Hubble Space
Telescope Ultra Deep Field (UDF) are generally exponential,
whereas the major axis profiles in irregular disk galaxies, called
clump-clusters in our previous studies, are clearly not. Here we
show that the deprojected positions of star-forming clumps in both
galaxy types are exponential, as are the deprojected luminosity
profiles of the total emissions from these clumps. These
exponentials are the same for both types when normalized to the
outer isophotal radii.  The results imply that clumps form or
accrete in exponential radial distributions, and when they
disperse they form smooth exponential disks. The exponential scale
lengths for UDF spirals average $\sim1.5$ kpc for a standard
cosmology. This length is smaller than the average for local
spirals by a factor of $\sim2$. Selection effects that may account
for this size difference among spirals are discussed. Regardless
of these effects, the mere existence of small UDF galaxies with
grand-design spiral arms differs significantly from the situation
in local fields, where equally small disks are usually dwarf
Irregulars that rarely have spiral arms. Spiral arms require a
disk mass comparable to the halo mass in the visible region --
something local spirals have but local dwarfs Irregulars do not.
Our UDF result then implies that galaxy disks grow from the inside
out, starting with a dense halo and dense disk that can form
spiral arms, and then adding lower density halo and disk material
over time. Bars that form early in such small, dense, gas-rich
disks should disperse more quickly than bars that form later in
fully developed disks.
\end{abstract}


\section{Introduction}
The origin of exponential disks has been difficult to observe
directly because the earliest examples of clear spiral galaxies
already have them (O'Neil, Bothun, \& Impey 2000; P\'erez 2004;
Elmegreen, Elmegreen, \& Sheets 2004; Toft et al. 2005).
Exponential disks have also been difficult to model in cosmology
simulations (van den Bosch 2001; Robertson et al. 2004), although
models with high numerical resolution (Governato et al. 2004),
star formation feedback (Westera et al. 2002; Abadi et al. 2003),
or nuclear black-hole feedback (Robertson et al. 2005) have made
exponential disks.

In two previous surveys, we distinguished between spiral galaxy
disks, which have bulges, spiral arms, and approximately
exponential profiles, and highly irregular disks, which we called
clump cluster galaxies (Elmegreen, Elmegreen \& Sheets 2004;
Elmegreen et al. 2005b). The association of clump clusters with
disks was made on the basis of models for the distribution of
axial ratios (Elmegreen, Elmegreen, \& Hirst 2004a; Elmegreen \&
Elmegreen 2005; Elmegreen et al. 2005b).  Edge-on clump clusters
are apparently chain galaxies, which also have no central bulges
or exponential disks, and have similar clumps (Elmegreen,
Elmegreen, \& Hirst 2004a).  In a study of 10 clump-cluster
galaxies, the average clump mass was found to be $\sim6\times10^8$
M$_\odot$ and the average clump diameter was 1.8 kpc (Elmegreen \&
Elmegreen 2005).

Clump clusters, by definition, do not have central light
concentrations resembling bulges, nor do they have exponential
disks along any radial profile. Some clump clusters, however, have
nearly exponential disks in their average radial light profile,
even though the major axis profile is irregular. Generally, the
elliptical contours for clump cluster galaxies cannot be
determined automatically by the IRAF routine {\it ellipse} because
the galaxy disks are too irregular. Still, we can use a constant
ellipticity and position angle to determine the average radial
profile, and obtain the values of the center, ellipticity and
position angle from the outer $2\sigma$ contour, where $\sigma$ is
the image noise level. Three examples of intensity profiles are
shown in Figure \ref{fig:profile}, along with the galaxy images.
On the left is a spiral with exponential profiles for both the
major axis (top panel) and the ellipse fits. In the central column
is a clump cluster with a typically irregular major axis profile
but a somewhat exponential elliptically-averaged profile. On the
right is another clump cluster with no exponential at all.

Exponential light profiles may be the result of cosmological
collapse during galaxy formation (Freeman 1970; and see references
above) or the result of radial flows in viscous, star-forming
disks (e.g., Lin \& Pringle 1987; Yoshii \& Sommer-Larsen 1989;
Zhang \& Wyse 2000; Ferguson \& Clarke 2001).  In the collapse
simulations, the gas may be accreted smoothly from the halo
(Murali et al. 2002; Westera et al. 2002; Sommer-Larsen, G\"otz,
\& Portinari 2003; Keres et al. 2005) or erratically from
cannibalized dwarf companions (Walker, Mihos, \& Hernquist 1996;
Abadi et al. 2003) and intergalactic clouds (Brook et al. 2004).

In the viscous models, the inner regions tend to move in, making a
central concentration, while the outer regions tend to spread out,
making an exponential in the outer disk.  The time scale for the
development of the outer exponential can be long, comparable to
the star formation time, if the initial disk is significantly
flatter than exponential or if it is highly irregular.  The
presence of exponential disks in local dwarf irregular galaxies
(Hunter \& Elmegreen 2005), where there is little shear for
viscous effects, and the early appearance of exponentials in
high-redshift spirals, suggest that this profile is related to
galaxy assembly, not shear.

In order to understand star formation in the disks of young
galaxies, we have been studying the properties of the bright blue
clumps which dominate the appearance of clump clusters (Elmegreen
\& Elmegreen 2005). Clump masses can be as large as $10^9$
M$_\odot$.  Most of these clumps have average densities only
slightly larger than the critical tidal densities in their disks,
so they should eventually disperse. Many also have tails or other
structures which make them look like they are in the process of
dispersing. As these galaxies also look very young (van den Bergh
et al. 1996; Conselice et al. 2004), clump dispersal in
clump-cluster galaxies could offer a clue to the origin of
exponential disks.

Here we show that even though the radial light profiles of clump
cluster galaxies are not exponential, the average radial profiles
of the clump positions are exponential, as are the average radial
profiles of the integrated clump fluxes. Thus, these galaxies may
evolve into exponential disks by the dissolution of the observed
clumps. In that case, it may be concluded that cosmological
accretion promotes exponential disks whether the gas comes in
smoothly or in the form of giant clumps. If the accretion is
smooth, then the disk has to be highly turbulent so that
gravitational instabilities form clouds and stellar clumps (Immeli
et al. 2004) with a Jeans mass as large as a clump mass.

\section{Small Exponential Scale Lengths in UDF Spirals}

The spiral or disk-like galaxies in our morphological catalog of
the Hubble Space Telescope Ultra Deep Field (UDF; Beckwith et al.
2005) were fit to elliptical isophotes using IRAF (Elmegreen et
al. 2005b). We considered only objects larger than 10 pixels in
diameter on the i$_{775}$ image. Those which had exponential light
profiles, spiral arms, and central bulges were designated as
spirals (269 galaxies) and those which did not were designated as
clump clusters (178 of them). All UDF spirals have an exponential
disk with a central bulge, although most of these bulges are
small.

\begin{figure}\epsscale{0.8}
\plotone{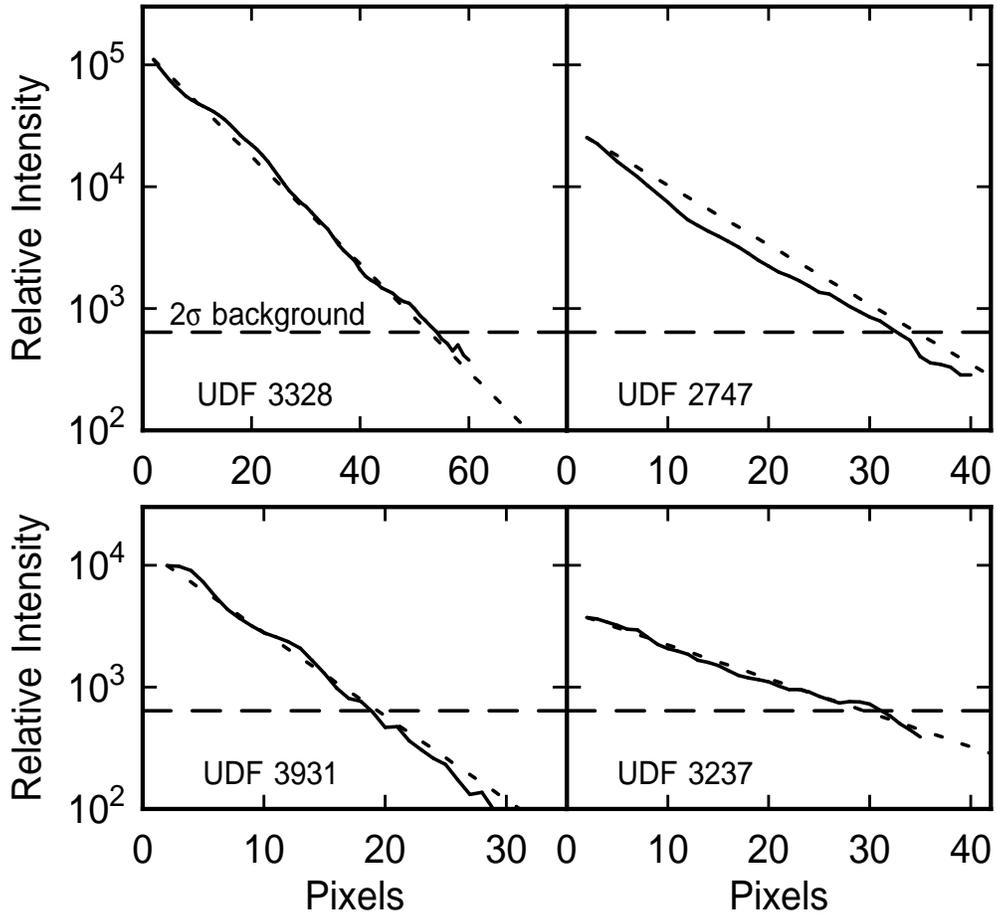}\caption{Four sample exponential disk fits using
the median technique.} \label{fig:samples}\end{figure}

For the spirals, we determined the exponential scale lengths in
pixels using the median value of the local slope on a log-linear
plot of ellipse-fit intensity versus radius in i$_{775}$ band.
This wavelength was chosen because it has the deepest UDF
exposures. The central pixel was not included in the ellipse fits.
The median technique avoids the small scale length of the bulge
region and possible scale length shifts in the far outer region.
The result is a good approximation to the scale length of the
main-disk exponential. Figure \ref{fig:samples} shows 4 examples
of the radial intensity profiles.  The dotted lines are the fits
and the dashed horizontal lines are the 2$\sigma$ background noise
levels. We do not discuss here double exponential disks or disk
truncations, such as those observed in local galaxies by Kregel \&
van der Kruit (2004), Erwin, Beckman, \& Pohlen (2005), and
others. Such profiles are present in our UDF sample and they have
also been noted for other high-redshift galaxies by P\'erez
(2004).

\begin{figure}\epsscale{0.8}
\plotone{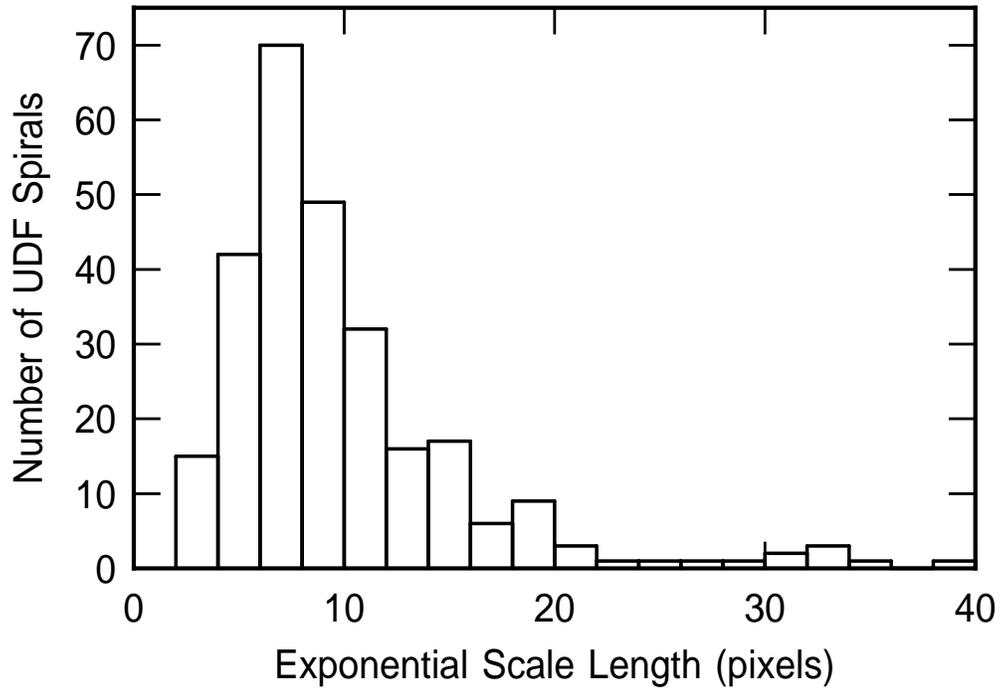}\caption{The distribution of exponential scale
lengths in spiral galaxies of the UDF.  The length is plotted in
units of a UDF pixel, which is 0.03 arcsec.}
\label{fig:rdisk}\end{figure}

\begin{figure}\epsscale{0.8}\plotone{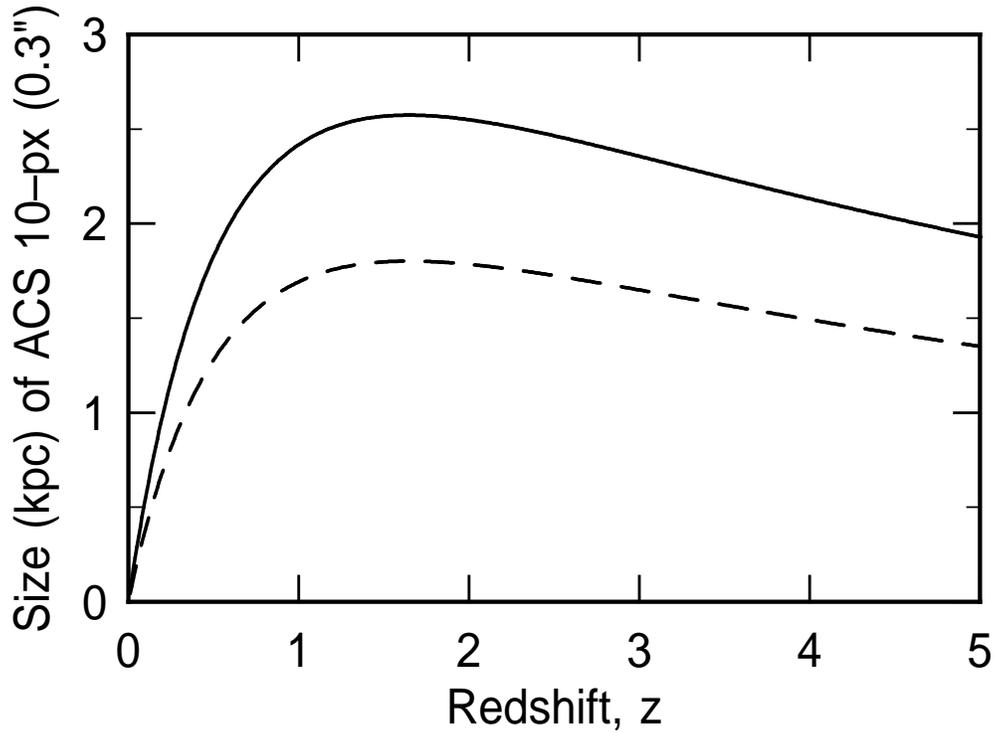}
\caption{The physical size of a region that subtends 10 pixels
(solid line) or 7 pixels (dashed line) of the ACS UDF field is
plotted as a function of cosmological redshift for the WMAP
cosmology. The conversion between angle and physical size is
relatively insensitive to $z$ in the range from 0.5 to 5, allowing
us to interpret the distribution of angular scale lengths shown in
the previous figure into an approximate distribution of physical
scale lengths.} \label{fig:chains6e}\end{figure}

Figure \ref{fig:rdisk} shows the distribution function of the
spiral galaxy scale lengths measured in pixels. Although the UDF
spirals probably span a wide range of redshifts, this distribution
function has physical meaning because the conversion between
angular size and physical size is relatively flat at redshifts
from 0.5 to 5. The conversion factor is shown in Figure
\ref{fig:chains6e} for the WMAP $\Lambda$CDM cosmology (Spergel et
al. 2003). The peak at $\sim7$ pixels in Figure \ref{fig:rdisk}
converts into 0.21 arc seconds for the ACS camera, in which case
Figure \ref{fig:chains6e} (dashed line) suggests that most spirals
have an exponential scale length of around 1.5 kpc, to within a
factor of $\sim1.5$.

This resultant average of $\sim1.5$ kpc for the UDF is smaller
than the average exponential scale length in local spiral
galaxies. van der Kruit (1987) obtained a local scale length
distribution with a peak at $3\pm2$ kpc for 42 galaxies having
known distances; the 31 spirals in his sample had an average scale
length of 4 kpc and the 11 dwarfs had an average scale length of 1
kpc. de Jong (1996a) observed B and K-band scale lengths somewhat
uniformly distributed between 1 and 7 kpc for 86 local galaxies,
but the volume-corrected distribution, which included substantial
corrections for missing small galaxies, peaked at 1.5 kpc and
extended from 0.8-11 kpc. Courteau (1996) found a scale length
distribution for 290 local Sb-Sc galaxies that peaked at $\sim3.5$
kpc with a range between 1.5 and 6 kpc. McGaugh \& Bothun (1994)
obtained a similar range of exponential scale lengths (1.2-6.3
kpc) for low surface brightness galaxies, as did de Blok, van der
Hulst, \& Bothun (1995), who got a median of 3.2 kpc. For local
dwarf galaxies, Hunter \& Elmegreen (2005) obtained an average
V-band scale length of 1 kpc for 94 galaxies of type Im, and 1.7
kpc for 18 galaxies of type Sm. Similar sizes were obtained in a
survey of 171 dwarf galaxies by Swaters \& Balcells (2002), who
found the distribution of scale lengths peaked at $\sim1$ kpc and
extended from approximately 0.2 to 4 kpc. Evidently, the UDF
spirals are systematically smaller than local spirals by a factor
of $\sim2$, and comparable in size to local spiral dwarfs (type
Sm). The local dwarf Irregulars (type Im) may be slightly smaller
than the UDF spirals.

The exponential scale length is a better measure of spiral galaxy
size than outer isophotal radius, which depends on the surface
brightness relative to the survey limit and varies with redshift
through cosmological dimming. The scale length may depend on
redshift through band shifting, as local disks have $\sim20$\%
larger B band scale lengths than K-band (de Jong 1996b; de Grijs
1998). Such color gradients typically come from star formation,
metallicity, and extinction gradients. The local-disk color
gradients are small, however, and if present at high z, would
increase the observed sizes, rather than decrease them, as the
i$_{775}$ band shifts to the rest frame blue.  The scale length
also depends only slightly on bulge-disk decomposition, becoming
smaller if the fitted exponential includes the bulge (Courteau
1996). This is not a consideration here because the median local
slope of the profile avoids the steeper slope of the bulge region,
as mentioned above, and because the UDF spirals do not have such
prominent bulges as local early-type galaxies.

de Jong (1996a) and Beijersbergen, de Blok, \& van der Hulst
(1999) noted that local galaxies have a slight correlation between
extrapolated central disk surface brightness and exponential scale
length in the sense that higher surface brightness galaxies are
intrinsically smaller (see also McGaugh \& Bothun 1994). The
difference in central surface brightness between local galaxies
with a scale length of 3.5 kpc (the average for the local-galaxy
surveys quoted above), and local galaxies with a scale length of
1.5 kpc (the average for our UDF spirals) corresponds to between 1
and 1.5 mag arcsec$^{-2}$ in rest frame B band.  Thus the small
sizes observed here for UDF spirals might be the result of an
observational bias toward high central surface brightness galaxies
in the UDF. Independent evidence for a $\sim25$\% loss of spirals
larger than 10 pixels below the UDF sensitivity limit was
presented elsewhere (Elmegreen, et al. 2005b), based partly on the
distribution of axial ratios. The magnitude of the selection
effect required for scale lengths seems too large to explain our
UDF distribution, however. To get the small scale lengths, we
would have to be selecting only UDF spirals that are intrinsically
brighter than the face-on average at high redshift by $1-1.5$ mag
arcsec$^{-2}$ in the center. This contrasts with our previous
study, where we estimated that the detection limit is only
$0.25-0.5$ mag arcsec$^{-2}$ brighter than the average surface
brightness. Larger galaxy losses at low surface brightness should
shift the peak in the axial ratio distribution to values lower
than $\sim0.5$ (minor to major axis ratio), where it is now. This
shift might be possible if internal dust removes such highly
inclined galaxies from the UDF survey. The galaxies that are in
our survey in fact have no correlation between central surface
brightness and scale length (the two quantities make a scatter
plot, which is not shown here). Thus the hypothesized large and
low-surface brightness galaxies have to be missing already from
our UDF survey. Nevertheless, it remains possible that UDF spirals
appear small because of surface brightness or other selection
effects.

A small size for high redshift galaxies has been noted before
(Lowenthal et al. 1997; Bouwens, Broadhurst, \& Silk 1998; Bouwens
\& Silk 2002; Ferguson, Dickinson, \& Giavalisco 2004; Papovich,
Dickinson, \& Giavalisco 2005; B\"ohm \& Ziegler 2005). Lyman
Break galaxies at higher redshift are even smaller (Baugh et al.
1998; Somerville, Primack \& Faber 2001). Small exponential scale
lengths ($\sim2$ kpc) were also found for $z\sim0.8$ barred
spirals in the Tadpole galaxy field (Elmegreen, Elmegreen \& Hirst
2004b). Disk scale lengths are expected to be smaller in galaxy
evolution models in proportion to the dark matter virial radius,
which varies as $H(z)^{-2/3}\sim\left(1+z\right)^{-1}$ for
constant mass, Hubble parameter $H(z)$, and large $z$ (Mo, Mau \&
White 1998; Bouwens, Broadhurst, \& Illingworth 2003; Ferguson,
Dickinson, \& Giavalisco 2004).

The scale length distribution function in Figure \ref{fig:rdisk}
is more revealing about galaxy size then the distribution of scale
length versus luminosity. Trujillo et al. (2005) found relatively
small scale lengths at high z for a given luminosity, but this may
be the case regardless of absolute length because high redshift
galaxies tend to have high surface brightnesses from high star
formation rates; then a fixed luminosity covers a smaller radius.
High intrinsic surface brightness is also expected from sampling
effects even without higher star formation rates because surface
brightness dimming hides the fainter populations.

The size evolution of a distribution of galaxies can differ from
the size evolution of any particular galaxy. If there is little
size evolution for particular galaxies, as suggested by Simard et
al. (1999) and Ravindranath et al. (2004), then most of our
spirals will become dwarfs or perhaps merge to form ellipticals.
Only the largest members of our sample could turn into spirals
today (e.g., Labb\'e et al. 2003).

One problem with this model of no size evolution is that the small
star-forming galaxies observed locally, which are typically types
Im, Sm, or BCD, do not have such clear spirals as the small UDF
spiral galaxies. The local dwarf disks are relatively thick (van
den Bergh 1988) and their stellar velocity dispersions relatively
large compared to their small rotation speeds, so they have little
ability to sustain spiral waves. Thus, distant spirals are
distinct from local dwarfs even at the same physical size.

To sustain a global spiral wave, the Toomre (1964) length, $2\pi
G\Sigma/\kappa^2$ (for epicyclic frequency $\kappa$ and total disk
surface density $\Sigma$) has to be comparable to the disk scale
length. The Toomre length is essentially the separation between
stellar spiral arms.  To generate a strong wave while avoiding
catastrophic collapse, the Toomre (1964) instability parameter,
$Q=\kappa c/\left(3.36 G \Sigma\right)$ for stellar velocity
dispersion $c$ has to be between 1 and 3.  If the disk scale
length is small for a galaxy that is morphologically spiral but
has a high surface brightness, then $\kappa$ has to be large.  The
stability parameter then requires $c$ to be fairly small, which is
possible if the disk mass is dominated by gas.  For the UDF
spirals, these constraints would be satisfied if the visible disks
are the inner gassy regions of fairly dense halo potentials, where
the density is about the same as in the inner regions of spiral
galaxies today. For a flat or solid body inner rotation curve,
$\kappa$ scales with the square root of the total enclosed
density.  At high redshift, galaxies should be not only smaller
but also denser in some inverse proportion (Mo, Mau \& White
1998), allowing tiny spirals to appear somewhat normal.

The Toomre length may also be viewed in another way. The ratio of
the Toomre length to the galaxy size is comparable to the ratio of
the disk mass to the total mass in the disk region. Thus, the
appearance of global spirals implies that disk masses are
comparable to halo masses out to the same radius. This is known to
be true for local spiral galaxies. Our observation of high
redshift spirals implies it is true there too, regardless of their
size. The situation is different for local dwarf irregulars,
however. There the ratio of disk mass to halo mass is relatively
small, i.e., they have relatively massive dark halos (e.g.,
Carignan \& Beaulieu 1989; Persic \& Salucci 1995). It follows
that the Toomre length is small compared to the galaxy size in
dwarfs, and so a stellar disturbance makes relatively small
epicycles and no global density waves.

These considerations lead us to believe that small UDF spirals are
dense like the inner regions of modern spirals. This should be the
case even if we select only the small members of a larger sample
because of surface brightness bias or other effects (see above).
The small spirals should grow to become large spirals by adding
low-density halo material to the periphery over time and by
filling out their disks with accreted gas and radially increasing
star formation (if they are not cannibalized first). This is the
commonly discussed inside-out formation model of galaxy disks but
now with an increasing scale length in addition to an increasing
overall size. A similar conclusion was reached by Papovich et al.
(2005) on the basis of an observed increase in half-light radius
and color gradient for $z\sim1$ compared to $z\sim2.3$ galaxies in
the Hubble Deep Field North. The color gradient in the $z\sim1$
galaxies was in the sense of a red core and a blue envelope, as if
star formation moved outward. Spiral arms were not so readily
resolved in that survey compared to the UDF, but some of the
galaxies studied by Papovich et al. are clearly spirals.  They
also point out that for such color gradients to exist, the major
merger activity has to be mostly finished by $z\sim1$, a
conclusion also reached by others for different reasons (e.g.,
Conselice, Blackburne \& Papovich 2005).

The higher density of high-redshift spirals has implications for
the formation and evolution of central bars.  Bar formation is
faster at higher density, and in a gas-rich young galaxy, also
more dissipative than usually considered in N-body simulations of
modern stellar disks.  Bar destruction with a central mass
concentration and gas torques should be faster at early times too,
for the same reasons. Moreover, if a fully-grown nuclear black
hole is present at this stage, then the disk mass fraction
represented by the black hole will be larger than it is for a
modern disk, thus promoting bar destruction by the degeneration of
x$_1$ orbits (e.g., Hasan, Pfenniger, \& Norman 1993). Modern
disks do not have the central mass concentration necessary for
such bar destruction (Shen \& Sellwood 2004; Athanassoula,
Lambert, \& Dehnen 2005), but very young disks could have. For
these reasons, it might be possible for young spiral disks that
are small, dense, and gas-rich to form and disperse relatively
small bars in their centers, thereby quickening the build up a
bulge from the disk.

\section{Clump Luminosity Fractions in UDF Disk Galaxies}

Galaxies at high redshift typically have giant blue clumps that
are associated with recent star formation.  In spirals, these
clumps are often in the arms, as they are in modern galaxies,
while in clump clusters, they are dispersed throughout the disk.
Equally large clumps are in chain galaxies (Cowie, Hu, \& Songaila
1995), high redshift-elliptical galaxies (Elmegreen, Elmegreen, \&
Ferguson 2005), tadpole galaxies (van den Bergh et al. 1996), and
double-clump galaxies (van den Bergh 2002).  A review of the
various morphologies seen in the UDF is in Elmegreen et al.
(2005b).

\begin{figure}\epsscale{0.6}\plotone{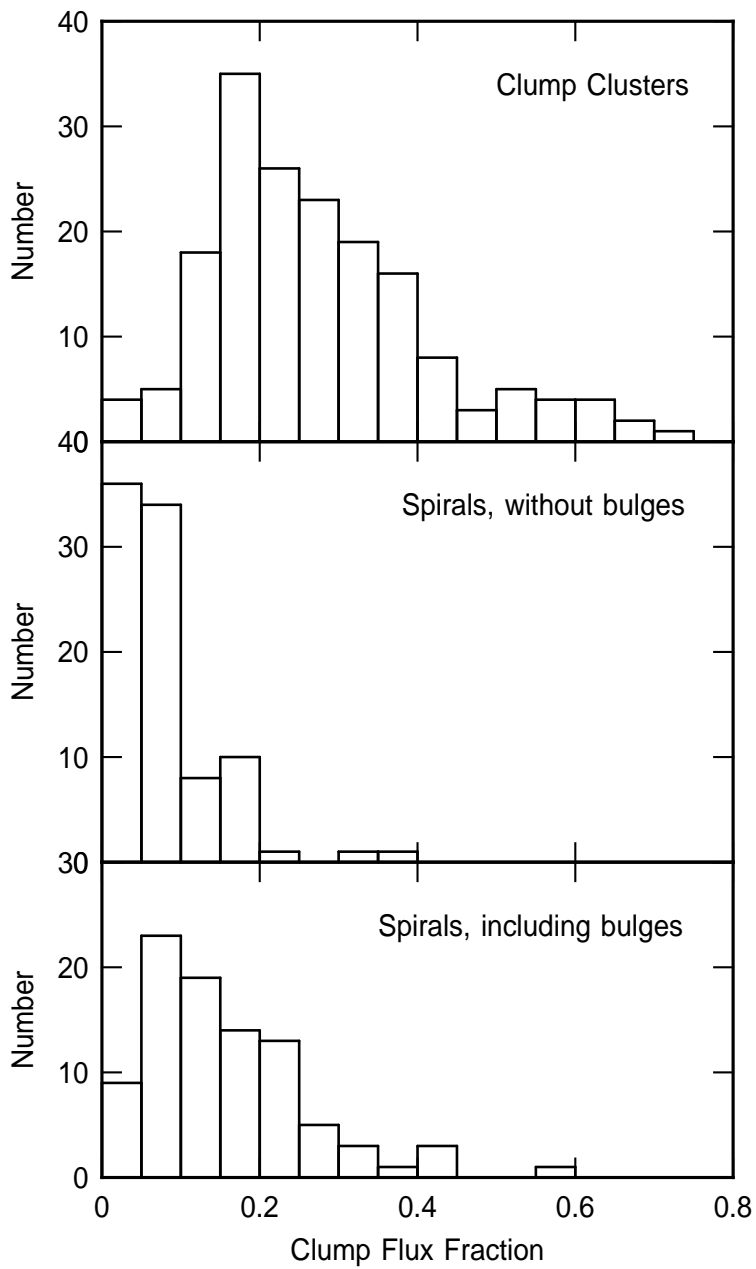}\caption{Histograms of the
fraction of the total galaxy light in the form of giant clumps,
determined at i$_{775}$.  The lower panel is for spirals where the
clumps include the central bulge; the middle panel is for spirals
without the bulges. The top panel is for clump cluster galaxies.
The clump clusters have $0.27\pm0.14$  of their measured i$_{775}$
flux in the form of giant clumps while the spirals have only
$0.080\pm0.065$ of their emission in such clumps, excluding the
bulges.} \label{fig:fraction}\end{figure}

Clump clusters differ from spiral galaxies not only in the
regularity of the clump positions, but also in the fraction of the
total light that is in the form of clumps. Figure
\ref{fig:fraction} shows histograms of this fraction using
magnitudes at i$_{775}$. Magnitudes were determined in IRAF using
a central aperture of 3 pixels radius and a background subtraction
annulus of 5 pixels width with a separation of 2 pixels between
them. The lower two panels are for the 91 relatively face-on
spiral galaxies in the UDF, with the bottom panel including the
central clumps or bulges in these galaxies and the middle panel
excluding the bulges. The bulges are generally more luminous than
the other clumps in the spirals, so the clump fraction is much
higher when the bulges are included. The top panel is for all 178
clump clusters, which have a very high fraction of their total
flux in the form of clumps.  The average flux fractions are
$0.27\pm0.14$ for clump clusters, $0.16\pm0.10$ for spirals
including their bulges, and $0.080\pm0.065$ for spirals not
including their bulges. Ten extreme cases of clump clusters were
studied in Elmegreen \& Elmegreen (2005) in which an average of
40\% of the i$_{775}$ flux was in the clumps.

The i$_{775}$ band corresponds to a different rest wavelength for
each galaxy, and the clump flux fraction is generally higher at
shorter wavelengths for local galaxies (i.e., star-forming regions
are blue). Thus one could interpret Figure \ref{fig:fraction} as
an indication that the clump clusters have systematically higher
redshifts than the spirals, with the shorter rest wavelengths more
clumpy. However, the large clump fraction at short wavelengths for
local galaxies is the result of a big difference between clump and
disk colors, the clumps being much bluer. This difference is not
as large at high redshift because the interclump stars are not as
old as they are locally. The average $B_{435}-V_{606}$ and
$V_{606}-i_{775}$ color differences between clump and interclump
emissions in the 10 clump cluster galaxies studied earlier are
$0.19\pm0.17$ mag and $0.14\pm 0.30$ mag, respectively, with the
clumps being bluer than the interclump regions.  Similarly, the
V$_{606}$-I$_{814}$ difference between the clump and surrounding
emission for the chain and spiral galaxies in the Tadpole field is
$0.2-0.3$ mag (Elmegreen, Elmegreen, \& Sheets 2004). In contrast,
the difference between clump and average colors in local spiral
galaxies is much greater. Local spiral galaxies have total
$B-V\sim0.7\pm0.1$ depending on Hubble type, and star forming
regions with $B-V\sim0$, so the clumps are bluer than the average
by $\sim0.7$ mag. The clump-interclump color difference is even
larger for local galaxies because the interclump emission is
redder than the average.

We consider in this paper that the difference in clump flux
fraction between spiral and clump cluster galaxies is a reasonable
indication that there is a real difference in the intrinsic
clumpiness of the two types of galaxies. A more quantitative
measure of disk clumpiness will require redshift information.

\section{Clump Position Distribution in UDF Disk Galaxies}

To study the relation between clump positions and exponential
disks, the (x,y) positions and B, V, i, and z magnitudes of all
the bright clumps in the low-inclination spiral and clump cluster
galaxies of the UDF were determined individually using IRAF.  This
represents 632 objects (including bulges) in the spirals and 904
objects in the clump clusters. The centers of all the galaxies
were measured also, using the middle of the central peak for the
spirals and the geometrically centered position of the outer
2$\sigma$ contour for the clump clusters. Deprojected
galactocentric radii were then determined from these positional
measurements, assuming a circular disk-like geometry with an
inclination from the axial ratio.

Because the galaxies span a wide range of redshifts, we normalized
the radii to the exponential scale length in the spirals, and to
the outer 2$\sigma$ isophotal radius for both the spirals and the
clump clusters.  This normalization also makes the rest wavelength
of the measurement relatively unimportant (it differs for each
galaxy) because the radial distribution of intensity is
effectively determined only for each galaxy separately and then
added together for all galaxies. Different rest wavelengths do not
affect the relative distribution of intensity inside each disk as
long as there are no systematic correlations between clump age and
radius.  In fact, clump B-V color has no measurable correlation
with deprojected radius for either galaxy type (not shown here).

\begin{figure}\epsscale{0.4}\plotone{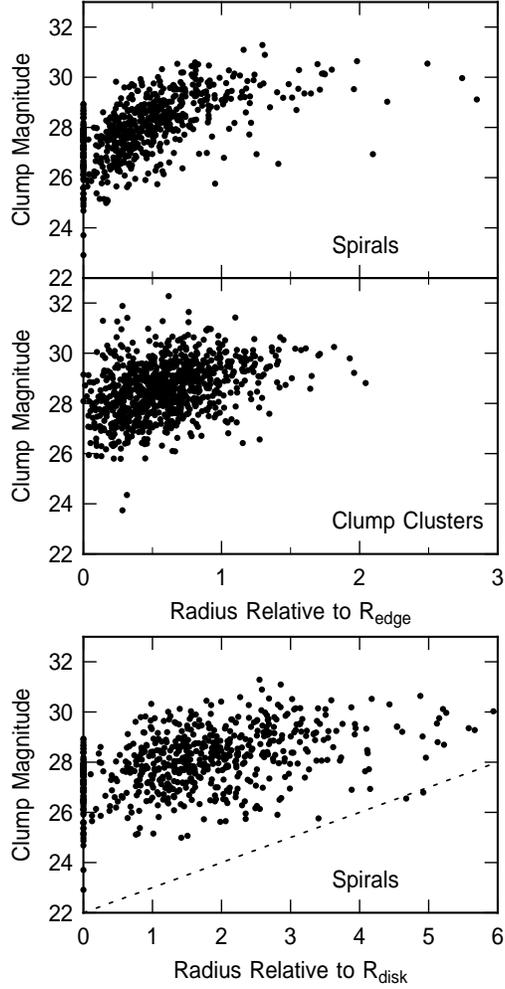}\caption{The distribution of
clump magnitudes in spiral and clump-cluster galaxies based on an
IRAF aperture 3 pixels in radius. The clumps get fainter with
deprojected distance from the galaxy center, nearly in proportion
to the background exponential disk for the case of spirals (lower
panel, dotted line), but also for the clump clusters where there
is relatively little background disk light. The top two panels
have galacto-centric radius normalized to the semi-major axis of
the outer isophotal contour, which is at a level of $2\sigma$ for
noise $\sigma$.  The bottom panel has the radius normalized to the
spiral scale length.} \label{fig:magdist}\end{figure}

Figure \ref{fig:magdist} plots the i$_{775}$ magnitudes of the
clumps versus the normalized galactocentric radii.  The points at
zero radius for the spirals are the bulges. The clumps get fainter
further from the center. The dotted line in the lower panel has
unit slope, corresponding to the magnitude-radius relation of each
underlying exponential disk. The distribution of points lies
parallel to this dotted line, suggesting a selection effect: we
can measure only the clumps that stand out sufficiently above the
background disk. Such an effect seems likely because star-forming
clumps in local galaxies have a wide range of luminosities with a
power-law distribution function dominated in number by the
smallest members. Thus, there may be countless small clumps that
were not measured. However, this should not be a problem for our
analysis because the local power laws have slopes of $\sim-2$ or
shallower for linear intervals of luminosity, and this means that
the integrated luminosity in all of the clumps in a particular
radial range is some fixed factor (less than $\sim5$) times the
luminosity of the largest member. The fact that there are no high
luminosity clumps in the outer regions (the lower envelopes in
Fig. \ref{fig:fraction} also increase with radius) means that
there really is an overall gradient in clump properties, although
we may not be seeing the true gradient with our selection effect.
This conclusion is supported by the observation that there is also
a tendency for the clumps to get fainter with galactocentric
radius in clump cluster galaxies. There are no significant
underlying disks in these cases and no likely systematic confusion
about clump definition and brightness at any radius.

\begin{figure}\epsscale{0.8}\plotone{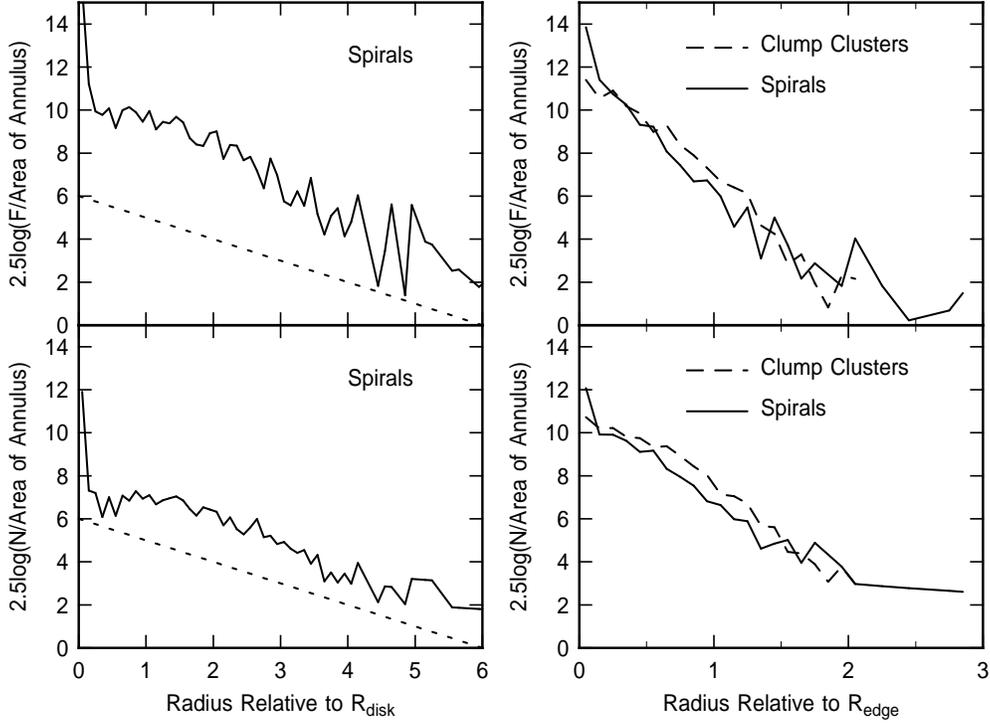}\caption{The distribution
with normalized galactocentric radius of the clump number density
(bottom) and the total clump flux density (top) in spiral and
clump-cluster galaxies. The normalization is with the disk scale
length on the left (only for spirals) and with the semi-major axis
of the outer $2\sigma$ isophotal contour on the right.  The
distribution of clumps follows the background exponential disk in
the case of spirals. The distributions for clump clusters and
spirals are about the same. This result implies that dispersing
clumps will convert the highly irregular disk of a clump-cluster
galaxy into one that is more like a spiral galaxy with an
exponential disk.}\label{fig:his}
\end{figure}

Figure \ref{fig:his} shows the radial distributions of clump
number density and clump flux density, converted to a magnitude
scale. The lower panels are for the clump number densities. These
were determined by summing the number of clumps in all galaxies of
each type that lie between radial intervals separated by 0.1 in
normalized units. On the left, the normalization of radius is with
the exponential scale length (for the spirals only) and on the
right, the normalization is with the radius of the outer 2$\sigma$
contour. The top panels are for the clump flux densities. These
were determined by summing the i$_{775}$ fluxes, $10^{-0.4m_I}$,
for all the clumps that lie between the same normalized radial
intervals. In both cases, the sums were divided by the areas of
the annuli corresponding to the radial interval and then converted
to a (negative) magnitude scale by the operation $2.5\log({\rm
sum/area})$. The dotted lines on the left have unit slope,
corresponding to the exponential disks in each spiral galaxy.  The
vertical scale in the figure has an arbitrary zero point with
higher densities toward the top.

The clump distributions in the spiral and clump cluster galaxies
show a remarkable similarity to each other even though the radial
profiles of the two types differ significantly in individual
cases.   This similarity is shown best by the agreement between
the dashed and solid lines in the right-hand panels. Evidently,
the clumps in clump cluster galaxies are mapping out an
exponential disk that is just like the exponential in spiral
galaxies.  This implies that if the clumps in clump cluster
galaxies are blended together, by dispersal of the associated
giant star complexes, for example, then an exponential disk will
result that is essentially the same as the disk in a spiral
galaxy. This similarity does not extend to the bulges, however:
the average density of clump fluxes at the centers of spirals is
higher than it is at the centers of clump clusters by 2 magnitudes
(a factor of 6 in the top right panel).  The bulges can be seen
again as the sharp rise at zero radius in the left-hand panels.

The correlation between clump flux and normalized galactocentric
radius (Fig. \ref{fig:magdist}) implies that the radial
distribution of average clump flux density (top of Fig.
\ref{fig:his}) is slightly steeper than the radial distribution of
clump number density (bottom of Fig. \ref{fig:his}). For radii
between $1.5R_{disk}$ and $5R_{disk}$, the slope of the
distribution on the left-hand plot of number density is $-1.3$ and
the slope of the distribution on the left-hand plot of flux
density is $-1.8$. Thus the integrated flux from clumps in the
spiral galaxies has an exponential scale length equal to
$1/1.8=0.55$ times the scale length of the total light. Dispersing
the clumps will shorten the overall scale length slightly (in
proportional to the relative luminosity of the clumps compared to
the total disk).

\section{Conclusions}

The exponential scale lengths for spiral galaxies in the UDF
appear to be a factor of $\sim2$ smaller than for local spiral
galaxies. Selection effects could give this result but the
required high central surface brightness seems more extreme than
permitted by the distribution of the ratio of axes. Because the
presence of global spiral waves implies that stellar disks are
relatively massive compared to the total mass inside the disk
radius, the small spiral galaxies in the UDF are probably only the
dense inner regions of today's disks. Subsequent accretion at
lower density should build up both the disk and the halo over
time, increasing the scale length and the overall galaxy size.
Bars that form early in dense, gas-rich disks would be more prone
to orbit degeneration and dispersal into a bulge than bars which
form later in fully developed disks.  N-body simulations of bar
formation should be revised to include more gas and to use
smaller, denser initial disks.

Galaxies in the UDF, including the spirals, are highly clumped
into blue star-forming regions that contain up to $\sim10^9$
M$_\odot$ of stars $\sim1$ Gy old or younger. The most clumpy of
the disk galaxies are called clump-clusters. In the
low-inclination clump-cluster galaxies studied here, $0.27\pm0.14$
of the total i$_{775}$ flux comes from giant disk clumps, compared
to only $0.080\pm0.065$ for the spirals. In a previous paper, we
suggested that many of these clumps should be dispersing as a
result of tidal forces. Here we addressed the question of what may
happen to the clumps after they disperse.

The evidence suggests that the clumps in clump-cluster galaxies
disperse into a more uniform disk that has an exponential radial
light profile.  We found that the clumps in both spirals and
clump-clusters were distributed in an exponential fashion, and
that their total flux was also distributed as an exponential. Even
though clump cluster galaxies do not presently have an exponential
light profile in a smooth disk, their average clump distribution
is as exponential as the profile in a spiral galaxy.

\begin{figure}
\epsscale{0.7} \plotone{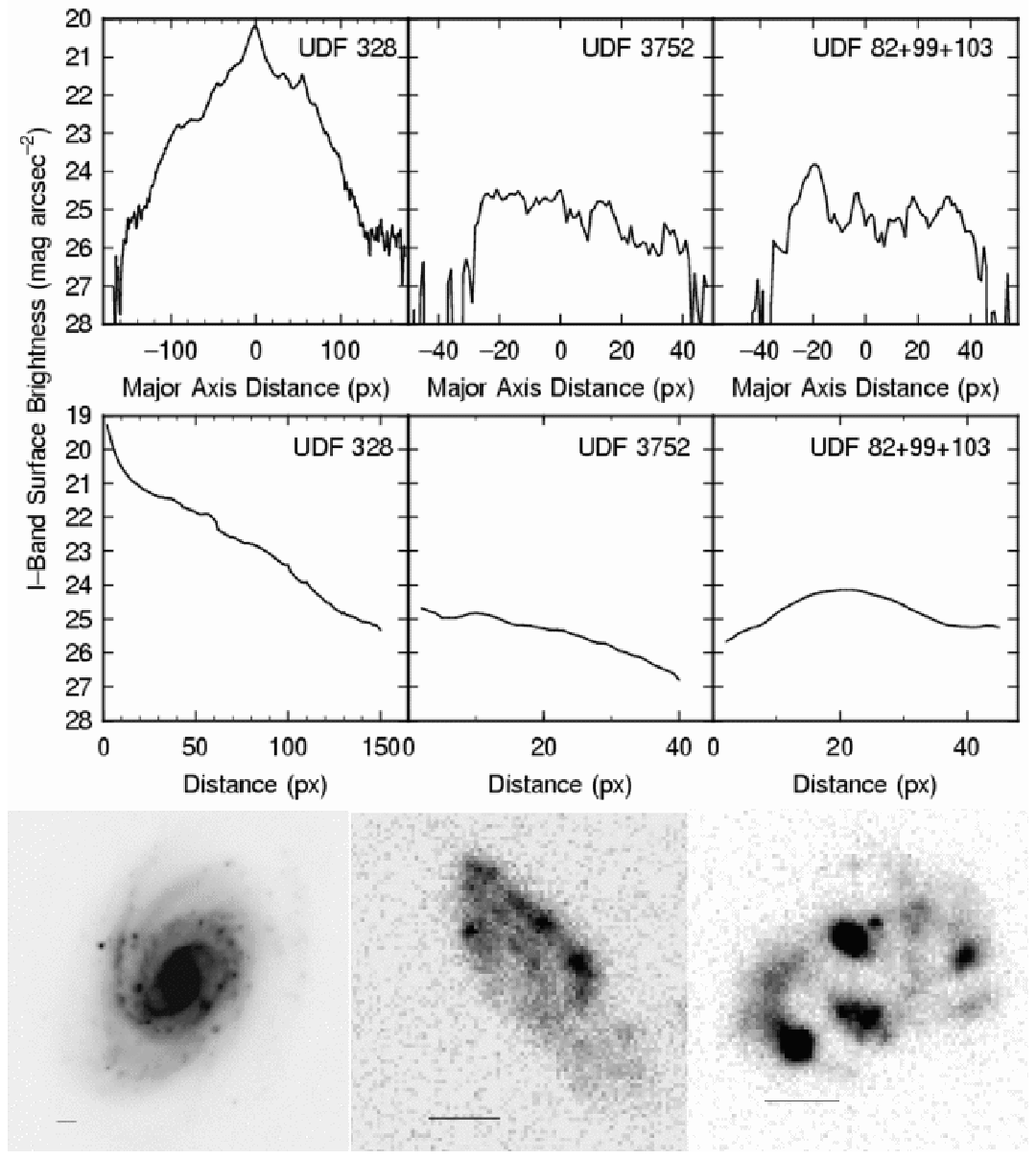} \caption{A spiral galaxy (UDF 328)
and two clump cluster galaxies (UDF 3752 in the middle column and
UDF82+99+103 on the right) are shown, with major axis profiles on
the top and elliptical-average radial profiles in the middle. UDF
328 has an exponential disk in both profiles, which is typical for
spirals, but the two clump-clusters, like most in their class,
have no exponential profiles along the major axes. The average
profiles for clump cluster galaxies may or may not be exponential.
UDF 3752 has an average profile that is exponential and UDF
82+99+103 has one that is not. The line in each grayscale image
represents 0.5 arcsec.} \label{fig:profile}\end{figure}

A similar conclusion was based on the near-exponential nature of
the average radial profiles of clump cluster galaxies made from
elliptical contours. While the major axis or single-cut profiles
are highly irregular, the average radial profiles, averaged over
projected ellipses in azimuth, are closer to exponential in many
cases (Fig. \ref{fig:profile}).

We also showed that spiral galaxy bulges stand out above the
exponential profile and are not part of it. Clump clusters
generally do not have such bulges. Thus it remains to be
determined whether clump clusters form primarily late Hubble-type
disks over time, or whether some of the clumps merge into a bulge,
as in models by Immeli et al. (2004).

Acknowledgments: We are grateful to the referee for comments that
improved the presentation of this paper. B.G.E. is supported by
the National Science Foundation through grant AST-0205097.

\clearpage

\newpage

\end{document}